\documentclass[a4paper]{jpconf}

\usepackage{amsmath}
\usepackage{amssymb}
\usepackage{epsfig}
\usepackage{graphicx,color}
\usepackage[sort&compress]{natbib}

\bibpunct{[}{]}{,}{n}{}{,}

\allowdisplaybreaks

\newcommand{\bra}[1]{\ensuremath{\langle #1 |}}   
\newcommand{\ket}[1]{\ensuremath{| #1 \rangle}}   
\newcommand{\sprod}[2]{\ensuremath{\left\langle #1 |%
                     #2 \right\rangle}}  

\renewcommand{\vec}[1]{{\mathbf{#1}}}

\newcommand{\arxiv}{}

\begin{document}

\title{The GSI anomaly}
\author{Hendrik Kienert$^1$, Joachim Kopp$^2$,
        Manfred Lindner$^3$,\\ Alexander Merle$^4$}
\address{Max--Planck--Institut f\"{u}r Kernphysik,
         Postfach 10 39 80, 69029 Heidelberg, Germany}
\ead{$^1$lars.hendrik.kienert@mpi-hd.mpg.de,
     $^2$joachim.kopp@mpi-hd.mpg.de,
     $^3$lindner@mpi-hd.mpg.de,
     $^4$alexander.merle@mpi-hd.mpg.de}

\begin{abstract}
Recently, an experiment at GSI Darmstadt has observed oscillating decay rates
of heavy ions. Several controversial attempts have been made to explain this
effect in terms of neutrino mixing.
We briefly describe the experimental results, give an overview
of the literature, and show that the effect cannot be due to
neutrino mixing. If the effect survives, it could, however, be explained by
hypothetical internal excitations of the mother ions ($\sim 10^{-15}$~eV).
\end{abstract}

\ifx \arxiv \undefined
Recently, an experiment at GSI Darmstadt has reported a 99\% C.L.~evidence for
an anomalous decay law in electron capture (EC) on hydrogen-like  ${}^{140}_{\
59}{\rm Pr}^{58+}$ and ${}^{142}_{\ 61}{\rm Pm}^{60+}$. Instead of the
pure exponential behaviour, a superimposed oscillation with a period of
$T \sim 7$~s was found.
\else
The accelerator facility at GSI Darmstadt can produce monoisotopic beams of
highly ionized heavy atoms and store them for extended periods of time in the
Experimental Storage Ring. There, an experiment has been performed in which
electron capture (EC) decays of hydrogen-like ${}^{140}_{\ 59}{\rm Pr}^{58+}$
and ${}^{142}_{\ 61}{\rm Pm}^{60+}$ ions have been studied using time resolved
Schottky mass spectrometry~\cite{Litvinov:2008rk}.  This technique allows to
detect changes of the ions' revolution frequencies which occur upon EC decay.
For a small number of stored ions ($\leq 3$), Schottky mass spectrometry allows
for a measurement of the individual decay times. Repeated measurements have led
to the distributions shown in Fig.~\ref{fig:exp-results}. On top of the
expected exponential behavior, there is a superimposed oscillation with a
period of $T \sim 7$~s.  The experiment avoids many systematical errors because
it provides a quasi-continuous monitoring of the ions.
Statistical fluctuations are excluded as the origin of the effect at the 99\%
confidence level~\cite{Litvinov:2008rk}.
\fi

Several authors have attempted to relate the anomaly to neutrino
mixing~\cite{Ivanov:2008sd,Kleinert:2008ps,Faber:2008tu,Lipkin:2008ai,Lipkin:2008in},
but these attempts have been refuted~\cite{Giunti:2008ex,
Giunti:2008im,Peshkin:2008vk}. In the following, we will first argue why the
GSI experiment is for general reasons distinct from a neutrino oscillation
experiment, and then present a careful treatment of the problem using the
density matrix formalism and the wave packet technique.  We will show that the
GSI anomaly cannot originate from neutrino mixing, but could be explained by
hypothetical internal excitations of the mother ions ($\sim 10^{-15}$~eV). In
the end, we will give a critical overview of the existing literature on
the subject.

\ifx \arxiv \undefined
\else
\begin{figure}
  \begin{center}
    \includegraphics[width=7cm]{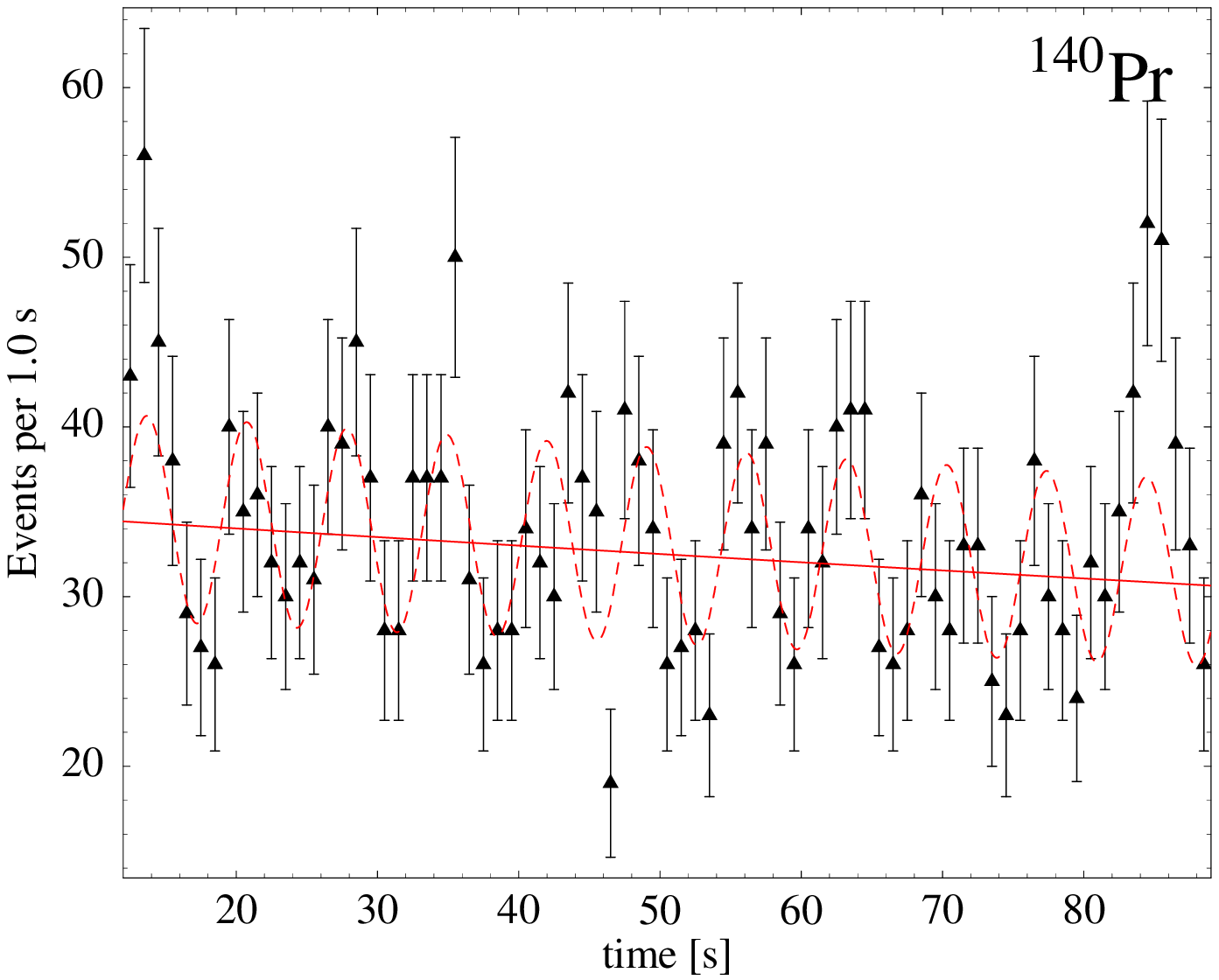} \hspace{0.8cm}
    \includegraphics[width=7cm]{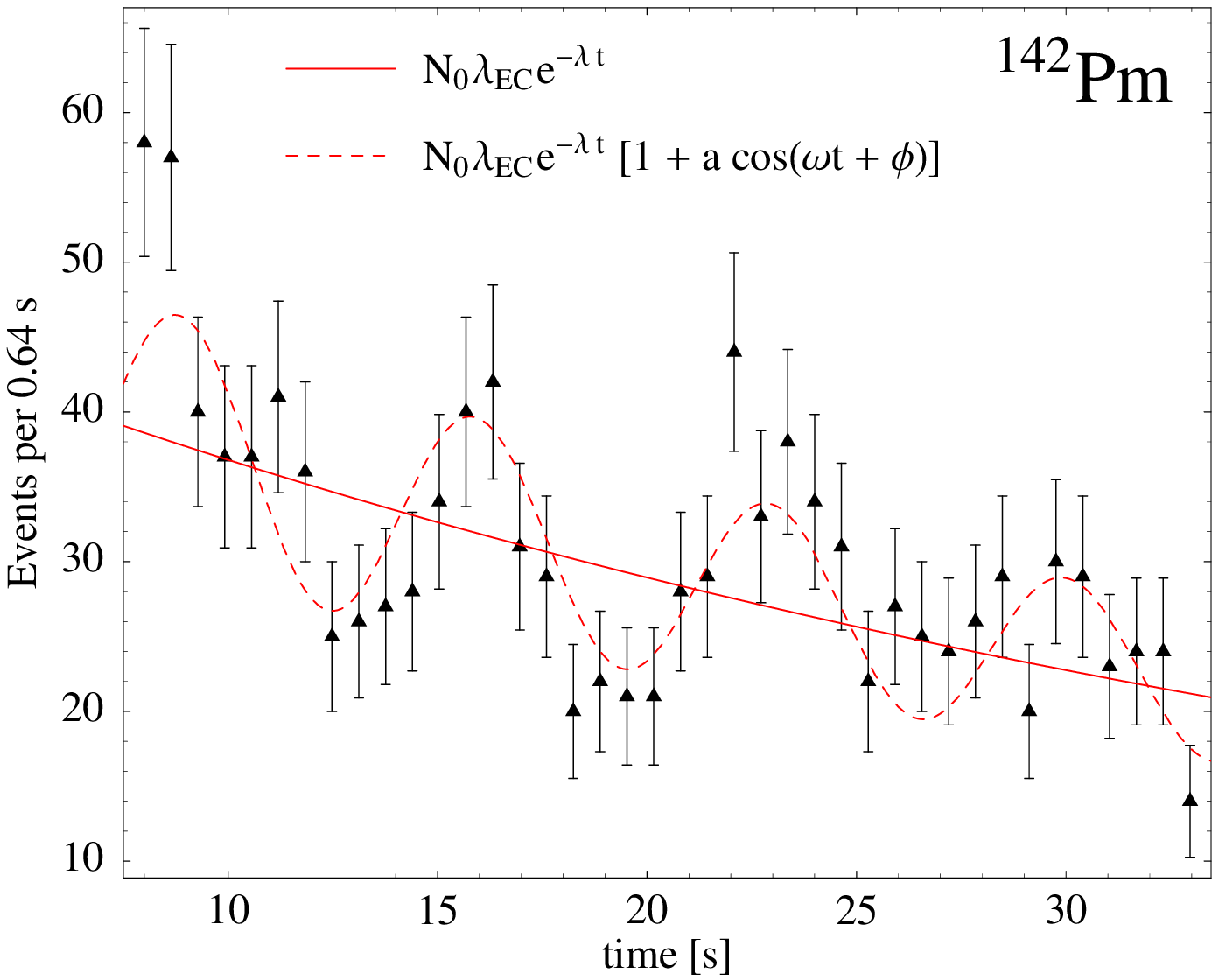}
  \end{center}
  \vspace{-0.3cm}
  \caption{Rate of EC decays of hydrogen-like ${}^{140}{\rm Pr}$
           and ${}^{142}{\rm Pm}$ ions as a function of the time after
           the injection into the storage ring. Data taken from the plots
           in~\cite{Litvinov:2008rk}.}
  \label{fig:exp-results}
\end{figure}
\fi

Let us start with a general argument why the GSI anomaly is distinct from neutrino oscillations, a fact which is not always correctly reported in the
literature~\cite{Walker:2008a}. Firstly, a Feynman diagram for neutrino
oscillations, e.g.\ for the $\nu_e$ to $\nu_e$ survival channel, has the
following form:
\begin{center}
  \includegraphics[height=2cm]{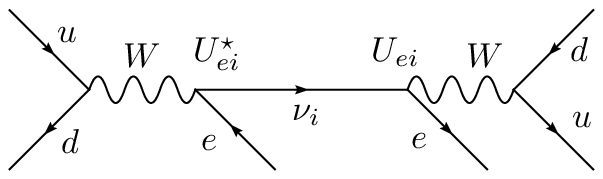}
\end{center}
Important are three different ingredients: The production process of the neutrino,
which is a weak interaction, ensures that the neutrino is produced in a flavour
eigenstate ($\nu_e$ in our example). The propagation of the neutrino has to be
described in terms of mass eigenstates because only those have definite
momenta (since the mass of a particle is explicitly included in its
propagator). The flavour eigenstate is a superposition of mass eigenstates,
$\ket{\nu_e} = \sum_i U_{ei}^* \ket{\nu_i}$, where $U_{ei}$ are the
corresponding elements of the leptonic mixing matrix. And finally, the
detection process involves another weak interaction, where the flavour of the
neutrino is measured {\it again}. Oscillations arise because the propagation
amplitudes of the three mass eigenstates are different, and hence their mixture
after some propagation distance is not the same as at the time of production.
The oscillation amplitude is given by a coherent sum,
\begin{align}
  A_{ee} = \sum_i |U_{ei}|^2 e^{-i p_i^\mu x_\mu},
  \label{eq:int1}
\end{align}
from which the standard quantum mechanical oscillation formula is easily obtained.

However, in the GSI experiment, the situation is completely different, since an
electron neutrino is produced, but there is {\it no second flavour measurement}. The
neutrino escapes undetected, as shown in the following Feynman diagram:
\begin{center}
 \includegraphics[height=2cm]{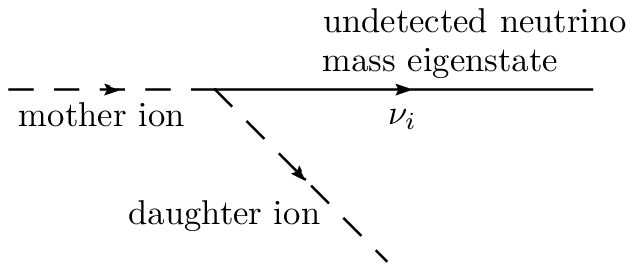}
\end{center}
Since Feynman diagrams describe transitions between states of definite energy
and momentum, the final state neutrino must be a mass eigenstate $\nu_i$ due to
energy-momentum conservation. In a hypothetical situation where the energies
and momenta of the mother and daughter ions are measured with
infinite precision, it would even be possible to tell from kinematics which
neutrino mass eigenstate has been produced.  Only this mass eigenstate,
i.e.~only one Feynman diagram, would contribute and the rate would be
proportional to $|U_{ei}|^2$, but there would be no oscillations.  However,
this is far from the real kinematical situation in the GSI setup.

Realistic energy and momentum uncertainties imply that it is not possible to
tell which mass eigenstate was produced, so that all of them have to be taken
into account, and must be treated as distinct final states. They contribute to
the total rate as an \emph{incoherent} sum of the sub-amplitudes, since each
mass eigenstate lives in its own Hilbert space.  The total rate is therefore
proportional to $\sum_i |U_{ei}|^2=1$, implying that in principle there
cannot be any mixing effects. Of course, this cannot change in a
quantum mechanical approximation of field theory.

Let us now discuss this in more detail, using the density matrix formalism for
a proper theoretical treatment of the detection process. We have to take into
account that the GSI detector is sensitive to the daughter ion, but not to the
neutrino. In the density matrix formalism, the probability for detecting a
particular state $\ket{n}$ in an experiment is given by $\tr (\hat{P} \rho)$,
where $\rho$ is the density matrix of the system, and $\hat{P}$ is the
projection operator onto $\ket{n}$. If one particle is not observed, the
projection is not onto one particular state, but rather onto a set of states
$\{ \ket{n,m} : n\ \text{fixed}, m = 0 \dots \infty \}$, where $n$ denotes the
degrees of freedom of the observed particles, and $m$ stands for those of the
unobserved particle.  In the GSI experiment, the detection of a daughter state
$\ket{\psi_{D,k}}$ is thus described by the operator
\begin{align}
  \hat{P}^{(k)} &= \sum_{j=1}^3 \int\!d^3p_\nu \,
             \ket{\psi_{D,k}; \nu_j, \vec{p}_\nu} \bra{\psi_{D,k}; \nu_j, \vec{p}_\nu}.
  \label{eq:proj}
\end{align}
The sum and the integral run over a complete set of neutrino mass
eigenstates $\ket{\nu_j}$ with momenta $\vec{p}_\nu$.
With the density matrix for the time-evolved mother state $\ket{\psi_M}$,
given by $\rho = \ket{\psi_M} \bra{\psi_M}$, the probability for
the observation of $\ket{\psi_{D,k}}$ becomes
\begin{align}
  \mathcal{P}_k &= \tr\Big[ \hat{P}^{(k)} \rho \Big]
    = \sum_{j=1}^3 \int\!d^3p_\nu \,
      \Big| \sprod{\psi_{D,k}; \nu_j, \vec{p}_\nu}{\psi_M} \Big|^2.
  \label{eq:Pk}
\end{align}
We see that the sum over neutrino states is incoherent. Therefore, if
$\mathcal{P}_k$ contains oscillatory interference terms, they cannot be due to
neutrino mixing, and would occur also in a hypothetical model with only one
neutrino flavour. All attempts to explain the GSI anomaly in terms of neutrino
mixing are thus foredoomed.

It is, however, imaginable that different components of the mother wave packet
$\ket{\psi_M}$ acquire relative phase differences during propagation.
If several such components could decay into the \emph{same} daughter state
$\ket{\psi_{D,k}; \nu_j, \vec{p}_\nu}$, they would induce interference terms in
$\mathcal{P}_k$. To see under which conditions such a mechanism could explain
the GSI anomaly, we will now compute the matrix element $\sprod{\psi_{D,k};
\nu_j, \vec{p}_\nu}{\psi_M}$ in the wave packet formalism (ref.~\cite{Beuthe:2001rc}
and references therein). We describe the states
$\ket{\psi_M}$ and $\ket{\psi_D}$ of the mother and daughter ions
by Gaussian wave packets
\begin{align}
  \psi_A(\vec{x}, t) &= \bigg( \frac{2\pi}{\sigma_A^2} \bigg)^{3/4} \!
    \int\!\frac{d^3p_A}{(2\pi)^3 \sqrt{2 E_A}} \,
    \exp\bigg[
          -\frac{(\vec{p}_A - \vec{p}_{0A})^2}{4\sigma_A^2}
          -i E_A \, (t - t_A) + i \vec{p}_A (\vec{x} - \vec{x}_A)
        \bigg] \, ,
\end{align}
where $A = M, D$. In our notation, $\vec{p}_{0M}$ and $\vec{p}_{0D}$ are the
central momenta of the wave packets, and $\sigma_M$, $\sigma_D$ are their
momentum space widths. $E_M$ and $E_D$ are related to $\vec{p}_M$ and
$\vec{p}_D$ by the relativistic energy-momentum relation.  $\psi_M$ is defined
such that, at the injection time $t_M$, its peak is located at $\vec{x}_M$.
Similarly, at the detection time $t_D$, the peak of $\psi_D$ is located at
$\vec{x}_D$.  The coordinate space Feynman rules yield for the amplitude of the
decay into $\ket{\psi_D; \nu_j,\vec{p}_\nu}$:
\begin{align}
  i \sprod{\psi_D; \nu_j, \vec{p}_\nu}{\psi_M} &=
    \int\!\frac{d^3p_M}{(2\pi)^3 \sqrt{2 E_M}}
    \int\!\frac{d^3p_D}{(2\pi)^3 \sqrt{2 E_D}}
    \int\!d^3x \, dt \,
    \bigg( \frac{2\pi}{\sigma_M \sigma_D} \bigg)^\frac{3}{2}
    U_{ej} \mathcal{M}^{EC}_j(E_M, E_D, E_\nu) \nonumber\\
  &\hspace{-3.0cm} \cdot
    \exp\bigg[
          -\frac{(\vec{p}_D - \vec{p}_{0D})^2}{4\sigma_D^2}
          + i E_D \, (t - t_D) - i \vec{p}_D (\vec{x} - \vec{x}_D)
        \bigg]                                 \nonumber\\
  &\hspace{-3.0cm} \cdot
    \exp\bigg[
          -\frac{(\vec{p}_M - \vec{p}_{0M})^2}{4\sigma_M^2}
          -i E_M \, (t - t_M) + i \vec{p}_M (\vec{x} - \vec{x}_M)
        \bigg]
    \exp\big[i E_{\nu,j} \, t - i \vec{p}_\nu \, \vec{x} \big],
  \label{eq:A-1}
\end{align}
with $\mathcal{M}^{EC}_j$ being the transition amplitude between plane
wave states, as computed in~\cite{Ivanov:2007pp,Bambynek:1977zz}.
\ifx \arxiv \undefined
  To evaluate the momentum integrals, we expand the complex phases
  to first order in $\vec{p}_M - \vec{p}_{0M}$ and $\vec{p}_D - \vec{p}_{0D}$
  (this approximation neglects the wave packet spreading~\cite{Beuthe:2001rc}).
  Moreover, we make use of the fact that the prefactors $1/\sqrt{2 E_{M,D}}$,
  as well as the matrix elements $\mathcal{M}^{EC}_j$ are almost constant
  over the width of the wave packets, and therefore can be replaced by
  their values at the central energies $E_{0M}$ resp.~$E_{0D}$. After
  these approximations, all integrals become Gaussian, and can be evaluated.
  We arrive at
\else
  To evaluate this expression, we are first going to compute the $p_M$ integral
  \begin{align}
    \int\!\frac{d^3p_M}{(2\pi)^3 \sqrt{2 E_M}} 
    \exp\bigg[
          -\frac{(\vec{p}_M - \vec{p}_{0M})^2}{4\sigma_M^2}
          - i E_M \, (t - t_M) + i \vec{p}_M (\vec{x} - \vec{x}_M) \bigg] \, .
  \end{align}
  We expand $E_M = (\vec{p}_M^2 + m_M^2)^{1/2}$ in the exponent up to first
  order in $\vec{p}_M - \vec{p}_{0M}$ (this approximation neglects wave packet
  spreading, as has been shown in~\cite{Beuthe:2001rc}). The prefactor
  $1/\sqrt{2 E_M}$ is assumed to vary slowly over the width of the wave packet,
  and will therefore be replaced by its value at $\vec{p}_{0M}$, which we
  denote by $1/\sqrt{2 E_{0M}}$. Similarly, we also neglect the energy
  dependence of the matrix element $\mathcal{M}^{EC}_j$. We then have to
  evaluate
  \begin{align}
    &\exp\big[-i E_{0M} \, (t - t_M) + i \vec{p}_{0M} (\vec{x} - \vec{x}_M) \big]
    \int\!\frac{d^3p_M}{(2\pi)^3 \sqrt{2 E_{0M}}} 
      \exp\bigg[
            -\frac{(\vec{p}_M - \vec{p}_{0M})^2}{4\sigma_M^2}
          \bigg]  \nonumber\\[0.2cm]
    &\hspace{5cm} \cdot
      \exp\big[i (\vec{p}_M - \vec{p}_{0M}) (\vec{x} - \vec{x}_M)
                    - i \vec{v}_{0M} (t - t_M) \big] \, ,
    \label{eq:wp-int}
  \end{align}
  where $\vec{v}_{0M} = \vec{p}_{0M} / E_{0M}$ is the group velocity of the
  wave packet. The result is
  \begin{align}
    \exp\big[-i E_{0M} \, (t - t_M) + i \vec{p}_{0M} (\vec{x} - \vec{x}_M) \big]
    \bigg( \frac{\sigma_M^2}{2 \pi} \bigg)^{3/2} \frac{2}{\sqrt{E_{0M}}}
    \exp\big[- \big(\vec{x} - \vec{x}_M - \vec{v}_{0M} (t - t_M) \big)^2 \sigma_M^2 \big] \, .
  \end{align}
  After evaluation of the $\vec{p}_D$ integral in a completely analogous way,
  eq.~\eqref{eq:A-1} becomes
  \begin{align}
    i \sprod{\psi_D; \nu_j, \vec{p}_\nu}{\psi_M} &=
      \frac{4}{\sqrt{E_{0M} E_{0D}}} \bigg(\frac{\sigma_M \sigma_D}{2 \pi} \bigg)^{\frac{3}{2}}
      \int\!d^3x \, dt \,
      U_{ej} \mathcal{M}^{EC}_j(E_{0M}, E_{0D}, E_\nu) \nonumber\\[0.2cm]
    &\hspace{-3.0cm} \cdot
      \exp\big[\!
            -i E_{0M} (t - t_M) + i \vec{p}_{0M} (\vec{x} - \vec{x}_M)
            - \big(\vec{x} - \vec{x}_M - \vec{v}_{0M} (t - t_M) \big)^2 \sigma_M^2
          \big] \nonumber\\[0.2cm]
    &\hspace{-3.0cm} \cdot
      \exp\big[\!
            -i E_{0D} (t - t_D) + i \vec{p}_{0D} (\vec{x} - \vec{x}_D)
            - \big(\vec{x} - \vec{x}_D - \vec{v}_{0D} (t - t_D) \big)^2 \sigma_D^2
          \big] \,
      \exp\big[i E_{\nu,j} \, t - i \vec{p}_\nu \, \vec{x} \big] .
  \end{align}
  We see that the $\vec{x}$-integral, as well as the subsequent $t$-integral,
  are Gaussian; a straightforward calculation thus leads to
\fi
\begin{align}
  i \sprod{\psi_D; \nu_j, \vec{p}_\nu}{\psi_M} &=
    \sqrt{\frac{2 \sigma_D \sigma_M}{\pi (\vec{v}_{0D} - \vec{v}_{0M})^2 E_{0M} E_{0D}}}
    \exp\big[\! -f_j \big] \exp\big[ i \phi_j \big] \,
    U_{ej} \mathcal{M}^{EC}_j\!(E_{M0}, E_{D0}, E_{\nu,j}) \, ,
  \label{eq:A-2}
\end{align}
with
\begin{align}
  f_j &= \frac{(E_j - \vec{p} \vec{v})^2}{4 \sigma_E^2} + \frac{\vec{p}^2}{4 \sigma_p^2}
          - \frac{\sigma_D^2 \sigma_M^2}{\sigma_p^2 (\vec{v}_{0D} - \vec{v}_{0M})^2}
            \big[ (\vec{y}_D - \vec{y}_M) \times (\vec{v}_{0D} - \vec{v}_{0M}) \big]^2 \, \label{eq:f-def} ,\\
  \phi_j &= \frac{(\vec{v}_{0D} - \vec{v}_{0M})(\vec{y}_D - \vec{y}_M)(E_j - \vec{p} \vec{v})}
            {(\vec{v}_{0D} - \vec{v}_{0M})^2} 
	  + \vec{p} \cdot \frac{\vec{y}_D \sigma_D^2 + \vec{y}_M \sigma_M^2}{\sigma_p^2},
                                                                              \nonumber\\
         &\hspace{2cm} + i E_{0M} \, t_M - i \vec{p}_{0M} \vec{x}_M
                       - i E_{0D} \, t_D + i \vec{p}_{0D} \vec{x}_D \,  \label{eq:phi-def}.
\end{align}
Here, we use the notation
\begin{align}
  E_j           = E_{0M} - E_{0D} - E_{\nu,j} \, , &\qquad\qquad
  \vec{p}       = \vec{p}_{0M} - \vec{p}_{0D} - \vec{p}_\nu \, ,  \nonumber\\[0.4cm]
  \sigma_E^2    = \frac{\sigma_D^2 \sigma_M^2 (\vec{v}_{0D} - \vec{v}_{0M})^2}
                        {\sigma_D^2 + \sigma_M^2} \, , &\qquad\qquad
  \sigma_p^2    = \sigma_D^2 + \sigma_M^2 \, ,                       \\[0.1cm]
  \vec{v}       = \frac{\vec{v}_{0D} \sigma_D^2 + \vec{v}_{0M} \sigma_M^2}
                        {\sigma_D^2 + \sigma_M^2} \, , &\qquad\qquad
  \vec{y}_{D,M} = \vec{x}_{D,M} - \vec{v}_{0D,0M} t_{D,M} \, .    \nonumber
  \label{eq:kin-def}
\end{align}
The group velocities of the wave packets are given by $\vec{v}_{0D,0M}
= \vec{p}_{0D,0M} / E_{0D,0M}$.
The real factor $\exp[-f_j]$ enforces sufficient overlap of the wave packets, but
is non-oscillatory for Gaussian wave packets. The complex phase factor
$\exp[i \phi_j]$ is oscillatory, but is irrelevant for the modulus of the matrix
element appearing in eq.~\eqref{eq:Pk}.

We will now construct a \emph{hypothetical} situation in which the GSI
oscillations \emph{can} be explained by a quantum mechanical interference
effect, namely by quantum beats of the mother ion.
This possibility has been pointed out previously
in~\cite{Giunti:2008ex,Giunti:2008im,Peshkin:2008vk}.  Let us assume that
the state of the mother ion is split into several sublevels $\ket{\psi_M^{(n)}}$,
and that, for some reason, the production process creates the mother ion
in a superposition
\begin{align}
  \ket{\psi_M} = \sum_n \alpha_n \ket{\psi_M^{(n)}} \, ,
\end{align}
where the coefficients $\alpha_n$ have to fulfill the normalization
condition $\sum_n |\alpha_n|^2 = 1$.
With this modification, eq.~\eqref{eq:A-2} turns into
\begin{align}
  i \sprod{\psi_D; \nu_j, \vec{p}_\nu}{\psi_M} \propto
    \sum_n \alpha_n
    \exp\big[ -f_j^{(n)} + i \phi_j^{(n)} \big] \, ,
  \label{eq:A-3}
\end{align}
where $f_j^{(n)}$ and $\phi_j^{(n)}$ are defined as in eqs.~\eqref{eq:f-def} and~\eqref{eq:phi-def},
but including an upper index ${(n)}$ for the quantities $E_{0M}$, $\vec{p}_{0M}$,
$\vec{v}_{0M}$, $E_j$, $\vec{p}$, $\vec{v}$, and $\vec{y}_{M}$. For
simplicity, we have neglected the $n$-dependence of the normalization
factors and of the matrix element. Typically, also the wave packet overlap factor $\exp[-f_j^{(n)}]$ will
be almost independent of $n$, so we can safely omit it in the following,
assuming it to be absorbed in the overall normalization constant.

Upon squaring $|\sprod{\psi_D; \nu_j, \vec{p}_\nu}{\psi_M}|$, we now
obtain interference terms proportional to $\exp[ i ( \phi_j^{(n)} -
\phi_j^{(m)} )]$.  To simplify these, let us go to the rest frame of the
daughter nucleus, in which $\vec{v}_{0D} = 0$ and $\vec{p}_{0D} = 0$. Moreover,
we will choose $\sigma_D = \sigma_M \equiv \sigma$, and we will
expand $\phi_j^{(n)} - \phi_j^{(m)}$ up to first order in the small quantities
\begin{align}
    \Delta E_{M0}^{(nm)} &\equiv E_{M0}^{(n)} - E_{M0}^{(m)}
      \simeq \xi \, \frac{\Delta m_{nm}^2}{2 E_{M0}^{(m)}} \, ,
    \qquad \\
    \Delta \vec{p}_{M0}^{(nm)} &\equiv \vec{p}_{M0}^{(n)} - \vec{p}_{M0}^{(m)}
      \simeq -(1 - \xi) \, \frac{\Delta m_{nm}^2 \, \vec{p}_{M0}^{(m)}}
                                {2 |\vec{p}_{M0}^{(m)}|^2} \, .
\end{align}
Here, $\xi$ is a real parameter that is determined by the details of
the production process. If we finally neglect terms of
$\mathcal{O}(\sigma/E_{M0}^{(n)})$, we find
\begin{align}
  |\sprod{\psi_D; \nu_j, \vec{p}_\nu}{\psi_M}|^2 &\propto
    \sum_{n,m} \alpha_n \alpha_m^*
    \exp\bigg[- i (\vec{x}_D - \vec{x}_M)
               \bigg( \frac{\Delta m_{nm}^2 \, \vec{p}_{M0}^{(m)}}
                           {2 |\vec{p}_{M0}^{(m)}|^2}
               \bigg) \bigg] \, .
\end{align}
Using the relation $\vec{x}_D - \vec{x}_M \simeq \vec{v}_{M0}^{(m)} \, (t_D -
t_M)$, which is a good approximation for sufficiently well localized wave
packets, the phase factor can equivalently be written as $\exp[-i (t_D - t_M)
\Delta m_{nm}^2 / 2 E_{M0}^{(m)}]$.  To explain the GSI oscillations with $T
\simeq 7$~s, one would require $\Delta m^2 \sim 2.2 \cdot 10^{-4} \ {\rm
eV}^2$, which corresponds to $|m^{(2)} - m^{(1)}| \sim 8.4 \cdot 10^{-16}$~eV. As has been pointed out in~\cite{Ivanov:2008zn}, there is no known
mechanism that could split up the ground state of the mother ion by such
a small amount, nor a known reason why the production process should
create a coherent superposition of the substates.

Before concluding, let us discuss why our results disagree with those of
several other authors, who have claimed that the GSI anomaly is a consequence
of neutrino mixing.  Ivanov, Reda, and Kienle~\cite{Ivanov:2008sd} perform a
calculation in which the amplitudes $\mathcal{A}_j = \mathcal{A}({}^{140}{\rm
Pr}^{58+} \rightarrow {}^{140}{\rm Ce}^{58+} + \nu_j)$, for $j = 1\dots3$,
receive different phase factors.  The appearance of these phase factors can be
traced back to the assumption of a finite domain for the time integral at the
Feynman vertex, and to the assumption of momentum non-conservation. However,
the authors sum the $\mathcal{A}_j$ \emph{coherently}, and thus obtain
oscillatory interference terms in the decay rate. To match the observed
oscillation period $T \sim 7$~s, a value of $\Delta m_{21}^2 \sim 2.22(3) \cdot
10^{-4} \ {\rm eV}^2$ is required for the solar mass squared difference, in
conflict with KamLAND results. In a later work~\cite{Ivanov:2008nb}, the
authors relate this discrepancy to loop-induced Coulomb interactions of the
neutrino. Moreover, they apply their formalism also to
$\beta^+$-decays~\cite{Ivanov:2008ig}.  The treatment of the detection process
in Refs.~\cite{Ivanov:2008sd,Ivanov:2008nb,Ivanov:2008ig} is in conflict with
our results, which show that the sum over the $\mathcal{A}_j$ should be
incoherent rather than coherent.  Similar arguments have been given previously
by Giunti~\cite{Giunti:2008ex,Giunti:2008im}, by Burkhardt et
al.~\cite{Burkhardt:2008ek}, and by Peshkin~\cite{Peshkin:2008vk}.  (Note that
Ivanov et al.~have replied to some of Giunti's remarks
in~\cite{Ivanov:2008xw,Ivanov:2008zn}.) Moreover, Giunti has shown another
problem, namely that the decay rate computed in~\cite{Ivanov:2008sd} does not
reduce to the Standard Model result if the neutrino masses are set to
zero~\cite{Giunti:2008im}.

Further explanation attempts for the GSI anomaly are due to
Faber~\cite{Faber:2008tu} and Lipkin~\cite{Lipkin:2008ai,Lipkin:2008in}. Both
authors employ various kinematical arguments and assumptions, which in both
cases yield relative phase differences for the $\mathcal{A}_j$. Faber and Lipkin
also sum the amplitudes \emph{coherently}, in conflict with
refs.~\cite{Giunti:2008ex,Giunti:2008im,Peshkin:2008vk,Burkhardt:2008ek} and
with our discussion.

Finally, Kleinert and Kienle propose an explanation of the GSI anomaly in terms
of a ``neutrino-pulsating vacuum''~\cite{Kleinert:2008ps}. The authors
interpret the $\nu_e$ emission in EC decay as the absorption of a negative
energy $\bar{\nu}_e$ from the Dirac sea forming the vacuum. They assume these
negative energy anti-neutrinos to undergo oscillations, and thus come to the
conclusion that the rate of EC decay should oscillate as well.  However, the
Dirac sea contains all three neutrino flavours, so that, due to unitarity, its
$\bar{\nu}_e$ charge remains constant over time. This remains true even if the
local density of $\bar{\nu}_e$ neutrino states should be modified by the
presence of an atomic nucleus, as alluded by Kleinert and Kienle.

Besides these theoretical works, two more experiments have been
performed investigating EC decays of \emph{stopped} ${}^{142}{\rm Pm}$ and
${}^{180}{\rm Re}$ \emph{atoms}~\cite{Vetter:2008ne,Faestermann:2008jt}
(see also~\cite{Litvinov:2008hf}). None of these experiments has found
any oscillatory signature, indicating that the anomaly, if physical,
must be related to the differing features of the GSI setup.

In conclusion, we have shown that neutrino mixing cannot cause the
oscillating electron capture decay rate of heavy ions that has been observed at
GSI. If the oscillations were confirmed, one must think of new physics, and we have
presented a possible, though exotic, mechanism which explains the
anomaly by hypothetical internal excitations of the mother ion ($\sim
10^{-15}$~eV).

We are grateful for many fruitful discussions with E.~Akhmedov, K.~Blaum,
F.~Bosch, A.~Ivanov, Y.~Litvinov, and A.~Smirnov. This work has been supported
by the Sonder\-forschungsbereich TR27 ``Neutrinos and Beyond'' der Deutschen
Forschungsgemeinschaft. HK and JK would like to acknowledge support from the
Studienstiftung des deutschen Volkes.

\begin{center}
  \rule{10cm}{0.25pt}
\end{center} 
\vspace{-0.7cm}
\bibliographystyle{apsrev}
\bibliography{gsi}

\end{document}